\documentclass{article}

\usepackage{arxiv}

\usepackage[utf8]{inputenc} 
\usepackage[T1]{fontenc}    
\usepackage[numbers]{natbib}

\usepackage{hyperref}       
\usepackage{url}            
\usepackage{booktabs}       
\usepackage{amsfonts}       
\usepackage{nicefrac}       
\usepackage{microtype}      
\usepackage{graphicx}
\usepackage{doi}

\usepackage{amsmath}
\usepackage{amssymb}
\usepackage{latexsym}
\usepackage{enumerate}
\usepackage{colortbl}
\usepackage{amsthm}
\usepackage{bm}
\usepackage{xcolor}
\usepackage{tikz, pgfplots, standalone} 
\pgfplotsset{compat=1.18} 
\usepackage{subcaption}  

\newtheorem{thm}{Theorem}

\newcommand{\ub}{\bm{u}}
\newcommand{\rhob}{\bm{\rho}}

\newcommand{\thetab}{\bm{\theta}}

\newcommand{\R}{\mathbb R}

\title{Modelling toroidal and cylindrical data via the trivariate wrapped Cauchy copula with non-uniform marginals}

\author{ \href{https://orcid.org/0009-0006-1372-7032}{\includegraphics[scale=0.06]{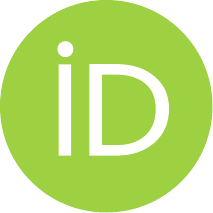}\hspace{1mm}Sophia Loizidou} \\
	University of Luxembourg\\
    Luxembourg \\
	\texttt{sophia.loizidou@uni.lu} \\
    \And
    \href{https://orcid.org/0000-0002-2290-8437}{\includegraphics[scale=0.06]{orcid.pdf}\hspace{1mm}Christophe Ley} \\
	University of Luxembourg\\
    Luxembourg \\
	\texttt{christophe.ley@uni.lu} \\
    \And
    \href{https://orcid.org/0000-0003-2648-6769}{\includegraphics[scale=0.06]{orcid.pdf}\hspace{1mm}Shogo Kato} \\
    Institute of Statistical Mathematics \\
    Japan \\
	\texttt{skato@ism.ac.jp} \\
	\And
	\href{https://orcid.org/0000-0003-0090-6235}{\includegraphics[scale=0.06]{orcid.pdf}\hspace{1mm}Kanti V. Mardia} \\
    University of Leeds \\
    United Kingdom \\
	\texttt{k.v.mardia@leeds.ac.uk} \\
}

\renewcommand{\shorttitle}{Modelling toroidal and cylindrical data with the trivariate wrapped Cauchy copula}

\hypersetup{
pdftitle={Modelling toroidal and cylindrical data via the trivariate wrapped Cauchy copula with non-uniform marginals},
pdfauthor={S. Loizidou, C. Ley, S. Kato, K. V. Mardia},
pdfkeywords={Angular data, copula, directional statistics, flexible modeling},
}

\begin{document}
\maketitle

\begin{abstract}
In this paper, we propose a new flexible family of distributions for data that consist of three angles, two angles and one linear component, or one angle and two linear components. To achieve this, we equip the recently proposed trivariate wrapped Cauchy copula with non-uniform marginals and develop a parameter estimation procedure.  We compare our model to its main competitors for analyzing trivariate data and provide some evidence of its advantages.  We illustrate our new model using toroidal data from protein bioinformatics of conformational angles, and cylindrical data from climate science related to buoy in the Adriatic Sea. The paper is motivated by these real trivariate datasets.
\end{abstract}


\section{Introduction}\label{sec:Intro}

Angular data are encountered in a wide range of disciplines, including environmental science (e.g., directions of wind or ocean waves), bioinformatics (such as dihedral angles in protein structures), zoology (tracking movement patterns of animals), medicine (e.g., circadian rhythms or hormone secretion timings), and the social or political sciences (for instance, the timing of crimes throughout the day); see, for example, \cite{LV19}. Analyzing such data requires particular attention due to their inherent circular nature, typically defined on the interval $[0, 2\pi)$ with endpoints identified. This circular topology renders many standard statistical methods for real-valued data inapplicable or inadequate, implying the need to develop appropriate directional statistics methods \citep{jammalamadaka_topics_2001, mardia2000}. In particular, this has led to a plethora of new probability distributions for data consisting of one angle (circular case), two angles (toroidal case), and one angle and one linear component (cylindrical case), see for instance \cite[Chapter~2]{LV19}, \cite{mardia_directional_2018} and \cite{pewsey_recent_2021}.

The situation is  different when the data consists of either three angles, two angles and one linear component, or one angle and two linear components, where only a few proposals exist and most of them have problems of tractability, complex parameter estimation procedures,
lack of practically usable random number generation or non-interpretable parameters (see Section~\ref{sec:compet}).  However, many such datasets {exist} which require an adequate statistical treatment. Two important examples are the following, which we will be analyzing with in this paper:
\begin{enumerate}
\item[1.] Protein data: Predicting the three-dimensional folding structure of a protein from its known one-dimensional amino acid structure is among the most important yet hardest scientific challenges, with impacts in drug development, vaccine design, disease mechanism understanding, human cell injection, and enzyme engineering, to cite but these. 
Demis Hassabis and John Jumper have been awarded the 2024 Nobel Prize in Chemistry for their deep learning based program AlphaFold which can  predict protein structures with very good accuracy \citep{senior2019protein}; see also \cite{mardia2025fisher}. 
The recent Nature Methods paper \citep{lane2023protein} concretely stated the need to complement AlphaFold single point prediction with an adequate statistical uncertainty treatment: ``distributions of conformations are the future of structural biology'' (see our Section~\ref{sec:protein} for details). So far most statistical advances, in particular on flexible and tractable probability distributions, have considered the two dihedral angles $\phi$ and $\psi$, and considered the torsion angle of the side chain $\omega$ to be fixed at either 0 or $\pi$ (which are the only two realistic values for this angle). One exception is  \cite{BMTFKH08} where  $\omega$ is taken as binary random variable with value either 0 or $\pi$ sampled using the hidden Markov model  though the distribution is discrete not continuous. In practice, this angle $\omega$ is often measured with some noise, hence a joint model for all three angles $\phi, \psi$ and $\omega$ is required.
\item[2.] Climate data: Many environmental agencies are collecting data on wave heights and {wave} directions in order to identify sea regimes. Such identification is highly relevant in climate-change for studies of, for instance, the drift of floating objects and oil spills \citep{huang2011wave}, coastal erosion \citep{pleskachevsky2009interaction}, and the design of off-shore structures \citep{faltinsen1993sea}; see \cite{lagona_hidden_2015} for further details. Typically, data analysis procedures use cylindrical distributions (e.g., the Abe--Ley distribution in \cite{lagona_hidden_2015}) to jointly model wave height and {wave} direction. A more scientifically strategic way is to add as additional variable the wind direction as it heavily influences waves {in confined spaces}. Quoting \cite{lagona_hidden_2015},``In wintertime, relevant events in the Adriatic Sea are typically generated by the southeastern Sirocco wind and the northern Bora wind". 
{That is, for such  cases, we should add wind direction  to the cylindrical data of wave height and wave  direction used in    \cite{lagona_hidden_2015} . Note that  their aim  is analyzing  a time series of cylindrical  data where the wind direction is not relevant. In our case we instead have three variables, wave height, wave direction and wind direction.}
\end{enumerate}


In the recent paper \cite{paper1}, we have proposed the trivariate wrapped Cauchy copula (TWCC) which precisely does satisfy the above-mentioned desiderata of tractability, good parameter estimation procedures,
 efficient random number generation and interpretable parameters. This copula can deal with three angles, yet it is itself not versatile enough because its marginals are uniform on the unit circle. Therefore, in this paper we propose an extension of the TWCC by adding general non-uniform marginal distributions to it, denoted TWCM for trivariate wrapped Cauchy copula with non-uniform marginals. 
Now, since the density of the uniform distribution on the circle exactly coincides with that of the uniform distribution on $[0,1]$, our construction implies that linear marginals are also admissible, in order to also cover cylindrical data. Our extended model in this paper thus not only covers toroidal data but also cylindrical data, unlike the toroidal competitor models from the literature.

The paper is organized as follows. In Section~\ref{sec:TWCC} we review the trivariate wrapped Cauchy copula as proposed in \cite{paper1}.
In Section~\ref{sec:adding marginals} we then introduce non-uniform marginals to the TWCC and hereby build the new flexible copula models TWCM, and we propose a parameter estimation procedure.
A thorough comparison with the main competitors in the existing literature is given in Section~\ref{sec:compet}, and in Section~\ref{sec:real_data} the two real datasets described above are investigated by means of our methodology.
Finally, Section~\ref{sec:conclusion} ends with a discussion and outlook on future research.


\section{The trivariate wrapped Cauchy copula}\label{sec:TWCC}

In this section we will briefly recall the definition and main properties of the trivariate wrapped Cauchy copula (TWCC)  of \cite{paper1}. 

For $\ub = \left( u_1, u_2, u_3 \right)'$ and $\rhob = \left( \rho_{12}, \rho_{13}, \rho_{23} \right)'$, the density of the TWCC is given by
\begin{align}
	\hspace{-0.1cm} t(\ub; \rhob) & =  c_2 \Bigl[ c_1 + 2 \left\{ \rho_{12} \cos (u_1 - u_2) + \rho_{13} \cos (u_1 - u_3) + \rho_{23} \cos (u_2 - u_3) \right\} \Bigr]^{-1},    \nonumber \\
	& \hspace{8cm} 0 \leq u_1,u_2, u_3 < 2\pi,  \label{eq:tri_density} 
\end{align}
where $\rho_{12},\rho_{13},\rho_{23} \in \mathbb{R} \setminus \{0\} $, $\rho_{12}\rho_{13} \rho_{23} >0$,
\begin{equation}
c_1 = \frac{\rho_{12} \rho_{13}}{ \rho_{23}} + \frac{\rho_{12} \rho_{23}}{\rho_{13}} + \frac{ \rho_{13} \rho_{23} }{\rho_{12}}, \label{eq:c1}
\end{equation}
and
\begin{equation} \label{eq:c2}
c_2 = \frac{1}{(2\pi)^3} \left\{ \left( \frac{\rho_{12} \rho_{13}}{ \rho_{23}} \right)^2 + \left( \frac{\rho_{12} \rho_{23}}{\rho_{13}} \right)^2 + \left(\frac{\rho_{13} \rho_{23}}{\rho_{12}} \right)^2 - 2 \rho_{12}^2 - 2 \rho_{13}^2 - 2 \rho_{23}^2 \right\}^{1/2}.
\end{equation}
Moreover, the parameters need to satisfy the condition  that there exists one of the permutations of $(1,2,3)$, $(i,j,k)$, such that 
\begin{equation}\label{conditions}
    |\rho_{j k}| < |\rho_{ij} \rho_{i k}| / ( |\rho_{ij}| + |\rho_{i k}|),\end{equation} where $\rho_{ji} = \rho_{ij}$ for $1 \leq i < j \leq 3$. 
If additionally they satisfy the identifiability constraint  \begin{equation}\label{identcond}
\rho_{12}\rho_{13}\rho_{23}=1,
\end{equation}
then the parameters of the TWCC are identifiable (see \cite{paper1} for the formal proof). Note that condition~\eqref{conditions} can be re-expressed under simpler form as $ \rho_{ij}^2 \rho_{i k}^2 >  (|\rho_{ij}| + |\rho_{i k}|)$ under \eqref{identcond}. We will equivalently refer to this distribution as TWCC
and TWCC$(\rhob)$ in order to specify its parameters.

In the {TWCC}, different functions of the parameters $\rho_{12}, \rho_{13}$ and $\rho_{23}$ control  the dependence between the $U_i$'s and the location of the modes, which have been well identified in \cite{paper1}.   An appealing aspect from  a tractability and computational viewpoint is the closed form of the density which does not include  integrals or infinite sums, unlike most of the existing distributions on the three-dimensional torus (see Section~\ref{sec:compet}). Moreover, all conditional distributions are members of well-known families, such as the wrapped Cauchy distributions on the circle and  the Kato--Pewsey distribution on the torus, and  the bivariate marginals are bivariate wrapped Cauchy-type {copulas}. A simple random number generation algorithm and an efficient parameter estimation procedure via  maximum likelihood estimation have been provided in \cite{paper1}.

\section{Adding non-uniform marginals to the TWCC} \label{sec:adding marginals}

Let $(U_1,U_2,U_3)'$ be a random vector which follows the {TWCC($\boldsymbol{\rho}$)}.
For each $i=1,2,3$, assume that either
\begin{itemize}
\item[-] $f_i$ is a density on the circle $[0,2\pi)$ and $F_i$ its distribution function with fixed and arbitrary origin, namely, $F_i (\theta) = \int_{c_i}^{\theta} f_i (x) d x$ with $c_i \in [0,2\pi )$; 
\item[-] $f_i$ is a density on (a subset of) the real line $\R$ and $F_i$ its distribution function. 
\end{itemize}
Define
$$
(\Theta_1,\Theta_2,\Theta_3)' = \left( F_1^{-1} \left( \frac{U_1}{2\pi} \right), F_2^{-1} \left( \frac{U_2}{2\pi} \right), F_3^{-1} \left( \frac{U_3}{2\pi} \right) \right)'.
$$
Then it follows that $(\Theta_1,\Theta_2,\Theta_3)'$ has the joint density
	\begin{equation}\label{eq:extended_density}
		 f(\thetab; \rhob) = (2\pi)^3 c_2 \Bigl( c_1 + 2 \sum_{\scriptsize 1 \leq i < j \leq 3} \rho_{ij} \cos [ 2 \pi \{ F_i(\theta_i) - F_j (\theta_j)\}] \Bigr)^{-1} \prod_{1 \leq k \leq 3} f_{k}(\theta_{k}),
	\end{equation} 
where $\thetab = (\theta_1,\theta_2,\theta_3)'$ and each $\theta_i$ either belongs to $[0,2\pi)$ or to (a subset of) $\R$. We refer to this extension of {TWCC($\pmb\rho$) with non-uniform marginals} as TWCM.
As is clear from the definition, the univariate marginal density of $\Theta_i$ is given by $f_i$ \ $(i=1,2,3)$. This yields a very flexible class of distributions with either 3 angular components, 2 angular and 1 linear components, 1 angular and 2 linear components, or even 3 linear components.

The reason why there is no distinction between the type of marginals lies in the simple fact that the uniform distribution on $[0,2\pi)$ bears by nature a double role as being linear as well as circular (since all points, and in particular the endpoints, share the same value). This unified viewpoint was  not taken up by the two papers \cite{johnson_angular-linear_1978} and \cite{WJ80}, of which the former introduced copulas for cylindrical data and the latter for toroidal data, nor in \cite{circulas} where the concept of copulas for circular data has been discussed in general. 

From the properties of TWCC($\rhob$), we can derive general results about the distribution of the extended TWCC, such as conditional distributions. In the next theorem, whose  proof is omitted (as it is similar to the proof of Theorems 2 and 3 of \cite{paper1}), we give the marginal and conditional distributions of the TWCM.
\label{sec: theorem_extended_TWCC}
\begin{thm}\label{theo_cop_mar}
	Let $(\Theta_1,\Theta_2,\Theta_3)'$ follow TWCM with density (\ref{eq:extended_density}).
	Then the following hold for $(\theta_1,\theta_2,\theta_3)'$ each either belonging to $[0,2\pi)$ or to (a subset of) $\R$:
	\begin{enumerate}
		\item[(i)] The univariate marginal distribution of $\Theta_i$ is $f_i(\theta_i)$.
		\item[(ii)] The bivariate marginal distribution of $(\Theta_i,\Theta_j)'$ is 
		$$
		f(\theta_i,\theta_j)=4\pi^2t_2(2 \pi  F_i(\theta_i), 2 \pi F_j(\theta_j); \phi_{ij})f_i(\theta_i)f_j(\theta_j),
		$$
				where, for $0 \leq u_i,u_j < 2\pi,$ we define 
                $$
t_2(u_i,u_j; \phi_{ij}) = \frac{1}{4\pi^2} \frac{ | 1-\phi_{ij}^2 | }{1+\phi_{ij}^2 - 2 \phi_{ij} \cos (u_i-u_j)} 
$$
with
\begin{equation}\label{thephi}
{\phi_{ij} =  \frac{1}{ 2 \rho_{ij} } \left\{ \frac{\rho_{i k} \rho_{j k}}{\rho_{ij}} - \frac{\rho_{ij} \rho_{i k}}{\rho_{j k}} - \frac{\rho_{ij} \rho_{j k} }{ \rho_{i k} } - (2\pi)^3 c_2 \right\} ,}
\end{equation} and $c_2$ is as in (\ref{eq:c2}).
		\item[(iii)] The bivariate conditional distribution of $(\Theta_i, \Theta_j)'$ given $\Theta_k=\theta_k$ is 
		$$
f(\theta_i,\theta_j|\theta_k)=4\pi^2 t_{2|1}(2\pi F_i(\theta_i),2\pi F_j(\theta_j)|2\pi F_k(\theta_k); \rhob)f_i(\theta_i)f_j(\theta_j),
		$$
		where, for  $0 \leq u_i,u_j,u_k < 2\pi,$ we have
$$ t_{2|1}(u_i,u_j|u_{k}; \rhob)  
		 =   2 \pi c_2 \, \Bigl[ c_1 + 2 \left\{ \rho_{ij} \cos (u_i - u_j) + \rho_{i k} \cos (u_i - u_{k}) + \rho_{j k} \cos (u_j - u_{k}) \right\} \Bigr]^{-1}.$$
        
		\item[(iv)] The univariate conditional distribution of $\Theta_i$ given $\Theta_j=\theta_j$ is 
		$$
		f(\theta_i|\theta_j)=2\pi t_{1|1}(2 \pi  F_i(\theta_i) | 2 \pi F_j(\theta_j); \phi_{ij})f_i(\theta_i),
		$$
		where, for $0 \leq u_i,u_j < 2\pi,$ we define $$t_{1|1}(u_i | u_j; \phi_{ij}) = \frac{1}{2\pi} \frac{ | 1-\phi_{ij}^2 | }{1+\phi_{ij}^2 - 2 \phi_{ij} \cos (u_i- u_j ) }$$
  with $\phi_{ij}$ given in \eqref{thephi}.
\item[(v)] The univariate conditional distribution of $\Theta_i$ given $(\Theta_j,\Theta_{k})'=(\theta_j, \theta_{k})'$ is
$$
f(\theta_i|\theta_j,\theta_k)=2\pi t_{1|2}(2\pi F_i(\theta_i)|2\pi F_j(\theta_j),2\pi F_k(\theta_k); \delta_{i|jk})f_i(\theta_i),
$$
where, for $0 \leq u_i,u_j,u_k < 2\pi,$ we define $$t_{1|2}(u_i | u_j,u_{k}; \eta_{i|jk} , \delta_{i|jk}) = \frac{1}{2 \pi} \frac{ | 1-\delta_{i|j k}^2 | }{1+\delta_{i|j k}^2 - 2 \delta_{i|j k} \cos (u_i- \eta_{i|j k})}$$
with $\eta_{i|j k} =  \arg (\phi_{i|j k})$ and $\delta_{i|j k} =|\phi_{i | j k}|$ for $\phi_{i|j k} = - \rho_{j k} (\rho_{ik}^{-1} e^{{\rm i} u_j} + \rho_{i j}^{-1} e^{{\rm i} u_{k}}) $.
		\end{enumerate}
\end{thm}

\

The results demonstrate that with our extended model all forms of regression analysis involving up to three angular and/or linear components are straightforward, which in a unified way covers what we described as our objective in Section~\ref{sec:Intro}. Random number generation from TWCM is also immediate by adding just the step 
$$
\Theta_ i = F_i^{-1} \left( \frac{U_i}{2\pi} \right),\, i=1,2,3,
$$
to the algorithm presented for the TWCC in \cite{paper1}.

The parameters of {the extended TWCC} as in~\eqref{eq:extended_density}  are estimated using MLE. When the marginals are uniform, the procedure from \cite{paper1} applies. In the case that the marginal distributions are not uniform, the maximum likelihood estimates of the parameters are calculated in a two-step approach. Let $\{ (\theta_{1m},\theta_{2m},\theta_{3m})' \}_{m=1}^{n}$ denote the sequence of toroidal observations from~\eqref{eq:extended_density}.
In the first step, the ML estimates of the parameters of the marginals are calculated, followed in the second step by estimating the parameters of the copula using $\{(u_{1m}, u_{2m}, u_{3m})' \}_{m=1}^{n}$ for
\begin{equation*}
    (u_{1m}, u_{2m}, u_{3m})' = \Bigl(2\pi F_1(\theta_{1m}; \hat{\vartheta}_1), 2\pi F_2(\theta_{2m}; \hat{\vartheta}_2), 2\pi F_3(\theta_{3m}; \hat{\vartheta}_3) \Bigr)',
\end{equation*}
where, for $i \in \{1,2,3\}$, $F_i$ represents the marginal density function of $\theta_i$ and $\hat{\vartheta}_i$ the parameters of $F_i$ obtained from the first step of the maximization. 
\cite{joe_multivariate_1997} refers to this as the method of inference functions for margins or IFM method. 
Efficiency and consistency of the estimates obtained with IFM compared to the estimates that can be obtained by performing one maximisation of the likelihood function can be found in \cite[Chapter~10]{joe_multivariate_1997} {and \cite[Chapter~5]{Joe15}}.

We conclude this section by briefly discussing one important choice of univariate marginals as in Theorem~\ref{theo_cop_mar}(i), namely when each $f_i$ is the wrapped Cauchy density
\begin{equation}
f_i (\theta) = \frac{1}{2\pi} \frac{1-\xi_i^2}{1+\xi_i^2-2\xi_i \cos (\theta-\mu_i)}, \quad 0 \leq \theta < 2\pi, \label{eq:wc_density}
\end{equation}
where $\mu_i$ is the location parameter and $\xi_i \in (-1,1)$ the concentration parameter. 
Assume that the origin of the distribution function $F_i$ is $c_i = \mu_i$.
In this case $F_i$ has the closed-form expression
$$
F_i(\theta_i) =  \frac{1}{\pi} \arctan \left( \frac{1+\xi_i}{1-\xi_i} \tan \frac{\theta_i-\mu_i}{2} \right) + I (\theta_i > \mu_i + \pi), \quad \mu_i < \theta_i < \mu_i + 2 \pi.
$$
Then, noting that $\cos (2 \arctan x) = (1-x^2)/(1+x^2)$ and $\sin (2 \arctan x) = 2x/(1+x^2) , \ x\in \mathbb{R},$ it can be shown that the density of (\ref{eq:extended_density}) can be expressed as
\begin{equation}
	f(\thetab; \rhob) = c_2 \prod_{i=1}^3 (1-\xi_i^2) \Biggl[ c_1 \prod_{i=1}^3 g_i (\theta_i) + 2 \sum_{\substack{1 \leq j< k \leq 3 \\ i \neq j,k}} \rho_{ij} { g_{k} (\theta_{k}) h_{ij} (\theta_i,\theta_j) } \Biggr]^{-1}, \label{eq:tri_cauchy}
\end{equation}
where {$g_i (\theta_i) = 1+\xi_i^2-2\xi_i \cos (\theta_i - \mu_i)$} and 
\begin{align*}
	\begin{split} 
h_{ij} (\theta_i,\theta_j) = \ & (1+\xi_i^2)(1+\xi_j^2) \cos (\theta_i-\mu_i) \cos (\theta_j - \mu_j)   \\
& + (1-\xi_i^2)(1-\xi_j^2) \sin (\theta_i-\mu_i) \sin (\theta_j - \mu_j) - 2 \xi_j (1+\xi_i^2) \cos (\theta_i-\mu_i) \\
& - 2 \xi_i (1+\xi_j^2) \cos (\theta_j-\mu_j) + 4 \xi_i \xi_j.
\end{split}
\end{align*}
Note that the density (\ref{eq:tri_cauchy}) does not involve any integrals.
As further nice properties derived from Theorem~\ref{theo_cop_mar}, the wrapped Cauchy copula with wrapped Cauchy marginals also has wrapped Cauchy conditionals for $\Theta_i$ given $\Theta_j=\theta_j$ and for $\Theta_i$ given $(\Theta_j,\Theta_{k})'=(\theta_j, \theta_{k})'$. Moreover, the bivariate marginal distribution of $(\Theta_i,\Theta_j)'$ is the bivariate wrapped Cauchy distribution of \cite{KP15} with density
  \begin{align*}
			f(\theta_i,\theta_j) = \tilde{\gamma} \, \{ \tilde{\gamma}_0 - \tilde{\gamma}_1
			\cos ( \theta_i - \mu_i ) - \tilde{\gamma}_2 \cos (\theta_j - \mu_j) & - \tilde{\gamma}_3 \cos (\theta_i -\mu_i) \cos (\theta_j - \mu_j) \\
   - & \tilde{\gamma}_4 \sin (\theta_i -\mu_i) \sin (\theta_j - \mu_j) \}^{-1},
		\end{align*}
		where $\tilde{\gamma}=(1-\rho_{ij}^2)(1-\xi_i^2)(1-\xi_j^2)/(4\pi^2)$, $\tilde{\gamma}_0 =
		(1+\rho_{ij}^2)(1+\xi_i^2)(1+\xi_j^2)-8 \rho_{ij} \xi_i \xi_j$, $\tilde{\gamma}_1= 2
		(1+\rho_{ij}^2) \xi_i (1+\xi_j^2)-4 \rho_{ij} (1+\xi_i^2) \xi_j$, $\tilde{\gamma}_2= 2
		(1+\rho_{ij}^2) (1+\xi_i^2) \xi_j-4 \rho_{ij} \xi_i (1+\xi_j^2)$,
		$\tilde{\gamma}_3=-4(1+\rho_{ij}^2)\xi_i \xi_j+2 \rho_{ij} (1+\xi_i^2) (1+\xi_j^2)$, and
		$\tilde{\gamma}_4= 2 \rho_{ij} (1-\xi_i^2)(1-\xi_j^2)$.
Random number generation from the distribution (\ref{eq:tri_cauchy}) can be carried out using the M\"obius transformation
$$
\Theta_ i = F_i^{-1} \left( \frac{U_i}{2\pi} \right) = \mu_i + 2 \arctan \left[ \frac{1-\xi_i}{1+\xi_i} \tan \left( \frac{U_i}{2} \right) \right].
$$

\section{Main competitors from the literature} \label{sec:compet}

Having discussed the main properties of our new TWCM, we can now proceed to a discussion of its main competitors from the literature. Bearing in mind that this is the first paper on trivariate copulas for {both} angular and cylindrical data, its competitors are in the angular (toroidal) case only. 

There is a $d$-dimensional toroidal copula model of importance given in \cite{kim_multivariate_2016}. Its construction is very general, but no code for its application is provided. The authors have performed a comparison of distinct models based on trivariate  angular protein data and have come to the conclusion that, in terms of AIC, the multivariate nonnegative trigonometric sums (MNNTS) model of \cite{fernandez-duran_modeling_2014}  is the best choice when $d=3$. The probability density function of the MNNTS
is given by
\begin{equation} \label {MNNTS}
f(\theta_1,\theta_2,\ldots,\theta_d) =\sum_{k_1=0}^{M_1} \sum_{k_2=0}^{M_2} \cdots \sum_{k_d=0}^{M_d} \sum_{m_1=0}^{M_1} \sum_{m_2=0}^{M_2} \cdots \sum_{m_d=0}^{M_d} c_{k_1 k_2 \cdots k_d} \bar{c}_{m_1 m_2 \cdots m_d} e^{\sum_{s=1}^d {\rm i}\left(k_s-m_s\right) \theta_s}
\end{equation}
where $c_{k_1 k_2 \cdots k_d}$ and $\bar{c}_{m_1 m_2 \cdots m_d}$ are the parameters with $\bar{c}$  the complex conjugate of $c$. Based on the findings from \cite{kim_multivariate_2016}, we will just focus on this model when analyzing protein data in Section~\ref{sec:protein}.

The paper \cite{kim_multivariate_2016} also considers the well-established multivariate von Mises distribution of \cite{mardia2008} which again does not do as well for that dataset though this model belongs to the exponential family  so inherits several desirable properties of  this family. 

There are other choices  such as  the multivariate Generalised von Mises (mGvM) distribution \citep{NFT2017}, the multivariate wrapped normal distribution \citep{Baba1981},
 the  multivariate  projected normal distribution 
\citep{mastrantonio_joint_2018}, and the inverse stereographic normal distribution \citep{Selvitella2019},  but these all have some limitations in comparison to our model such as either their normalizing constants are intractable  so the MLE  computations are not easy, they may be multimodal, or sampling is not clear-cut.  Hence, we have selected only the MNNTS as competitor in Section~\ref{sec:protein} for our protein data. 

{ 
\section{Practical applications} \label{sec:real_data}

\subsection{Protein data}\label{sec:protein}

 As mentioned in Section~\ref{sec:Intro}, an essential challenge in the prediction of the 3D structure of proteins is the ability to jointly model the conformational angles $\phi,\psi$ (dihedral angles) and $\omega$ (torsion angle of the side chain). 
 The main body of statistical research in this direction has so far concentrated on the two conformational angles $\phi,\psi$, which has already led to important contributions in the protein structure prediction problem, see for example \cite{BMTFKH08}, \cite{HBPFJH10}, \cite{GSMH19}, \cite{HMF12}, and Chapters 1 and 4 of \cite{LV19}. This is due to the fact that, theoretically, the torsion angle $\omega$ is either  0 or $\pi$ radians, corresponding to peptide planarity. However, \cite{berkholz2012nonplanar} have investigated this issue in detail and shown that peptide planarity is not a common feature for collected data. In their paper, they conduct an  exploratory data analysis of all three conformational angles. The authors from \cite{mardia2007}, \cite{mardia2008} and \cite{kim_multivariate_2016} have modelled three or more angles with their proposed distributions.   The main problem of statistical approaches to protein structure prediction is that they are computationally intensive (especially due to intricate normalizing constants), and we believe that our proposal, which is a faster method, can  make the statistical approach more feasible.

We now give some background to  protein structure. The building blocks of proteins are the amino acids, which consist of the backbone and the sidechains.
The backbone consists of the chemical bonds NH-C$\alpha$ and C$\alpha$-CO, where C$\alpha$ denotes the central Carbon atom.
These bonds can rotate around their axes, with $\phi$ denoting the NH-C$\alpha$  angle and $\psi$ the C$\alpha$-CO angle. 
These angles need to be studied as specific combinations of them allow the favourable hydrogen bonding patterns, while others can `result in clashes within the backbone or between adjacent sidechains' \citep{jacobsen_introduction_2023, Mardia13, Mardiachapter}. 
The $\omega$ angle denotes the N-C torsion angle, where C is the non-central carbon atom. 
As we have said before, theoretically the angle $\omega$ can only take the values $0$ and $\pi$, and in some of the research  work, this angle was  fixed at one of the two values \citep{AL22, HBPFJH10}. 
However, in practice this angle is often measured with some noise and using our {extended TWCC}, we shall model all three angles.
In bioinformatics, and as already mentioned, all three angles are studied in  \cite{berkholz2012nonplanar} and we will give some comments on how their exploratory work  is explained by our model.

For the present data analysis, we consider position 55 at 2000 randomly selected times in the molecular dynamic trajectory of the SARS-CoV-2 spike domain from \cite{genna_sars-cov-2_2020}.
The position occurs in $\alpha$-helix throughout the trajectory. DPPS \citep{kabsch_dictionary_1983} is used to compute the secondary structure and \cite{cock_biopython_2009} to verify the chains. As can be seen from Figure~\ref{fig:rose_plots}, the marginal distributions of the angles $\phi, \psi$ and $\omega$ for this dataset cannot well be modelled by the uniform distribution on the circle (as done in \cite{paper1}), so other distributions need to be explored. 
We combined the TWCM with wrapped Cauchy, cardioid, von Mises and Kato--Jones marginals. 
Of course, many other choices are possible, and one can also combine distinct marginals. Whatever marginals we choose, the parameters of our models are estimated using MLE (IFM) as explained in Section~\ref{sec:adding marginals} and a model comparison is carried out using the Akaike Information Criterion (AIC) and the Bayesian Information Criterion (BIC).

\begin{figure}[htbp]
	\begin{tabular}{ccc}
		\centering 
         \includegraphics[width = 15cm, height = 5cm]{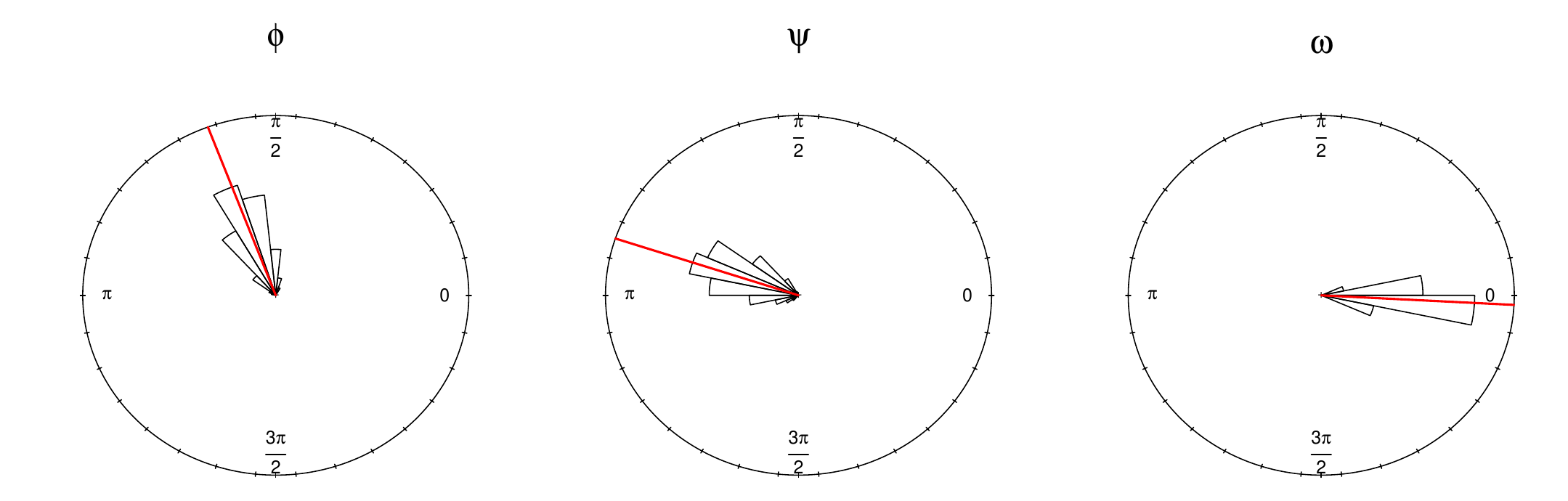}
\end{tabular}
\caption[]{\label{fig:rose_plots}Rose plots of the protein data for the three angles $\phi, \psi, \omega$. 
The marginals are not uniformly distributed on the circle. The {mean direction}, calculated using von Mises marginals, is plotted in red.}
\end{figure}

As conditional plots of two angles given one turn out to be informative in our setting, Figure \ref{fig:protein_contour_plots} shows the plots of the data of two of the angles given the third one. 
The values of the fixed angles in radians ($\phi = 1.93$, $\psi = 2.8$ and $\omega = 0$) are chosen to be the mode of the data, and only points that are within 0.1 radians of the selected value of the fixed angle are plotted.
In order to make the plots clearer, the range of values on each axis is not between $0$ and $2\pi$, like the traditional Ramachandran plot, but it is chosen such that both the contour plots and the points are visible (in summary, we have re-scaled the plots). 
For the values of $\omega$, most observations were around 0 and so the plot is translated from $[0, 2\pi$) to $[-\pi, \pi)$ in order to be able to make the range of values shown on the axis smaller and hence the presentation clearer.
The contour plots correspond to our copula density \eqref{eq:tri_density} with the marginal distributions being the von Mises distribution, whose density is given by
\begin{equation}\label{eq:dens_vonMises}
    f(\theta) = \frac{1}{2\pi I_0(\kappa)} \exp^{\kappa \cos(\theta-\mu)}, \quad \theta\in [0,2\pi),
\end{equation}
where $I_0(\kappa)$ denotes the modified Bessel function of the first kind and order 0. The von Mises marginals with their lighter tails lead to the best fit for this concentrated dataset, as measured by both  AIC and  BIC, see Table~\ref{tab:competitors}. 
The observed data points are plotted on top of the contours. 
This gives a visual indication of how good our estimated model fits this protein dataset.

For TWCM with von Mises marginals, the standard errors of the estimated parameters can be evaluated by means of bootstrap. 
Using $B=1000$ bootstrap samples, the maximum likelihood estimates of the parameters, with the standard error given in parenthesis, are $\hat{\rho}_{12} = 9.18 (4.99), \hat{\rho}_{13} = -1.17 (4.60), \hat{\rho}_{23} = -0.09 (4.41)$ for the copula parameters and $\hat{\mu}_1 = 1.93 (0.0043), \hat{\kappa}_1 = 27.6 (0.99), \hat{\mu}_2 = 2.82 (0.0055), \hat{\kappa}_2 = 17.3 (0.62), \hat{\mu}_3 = 6.23 (0.0024), \hat{\kappa}_3 = 84.4 (2.95)$, where $\hat{\mu}_i$ and $\hat{\kappa}_i$ denote the estimated values for $\mu$ and $\kappa$ of density \eqref{eq:dens_vonMises} corresponding to the marginal distribution of $\theta_i$ for $i\in\{1,2,3\}$. We attribute the relatively high standard errors of the copula parameters to the condition~\eqref{identcond} which links all three parameters.

Our findings regarding the angle $\omega$ confirm the exploratory analysis done by \cite{berkholz2012nonplanar}: the data points are strongly concentrated around the modal direction $\hat{\mu}_3$, but are not exactly equal to that value, as can also be visually appreciated from the plot in Figure~\ref{fig:rose_plots}.
The inherent tractability of our model allows biologists and bioinformaticians to quantify uncertainties, compute quantities of interest, and in particular our straightforward random number generation process enables them to quickly simulate data from our model, which is essential in their pipelines \citep{thygesen_efficient_2021}.

\begin{figure}[htbp]
	\begin{tabular}{ccc}
		\centering 

        \includegraphics[width = 7.3cm, height = 7cm]{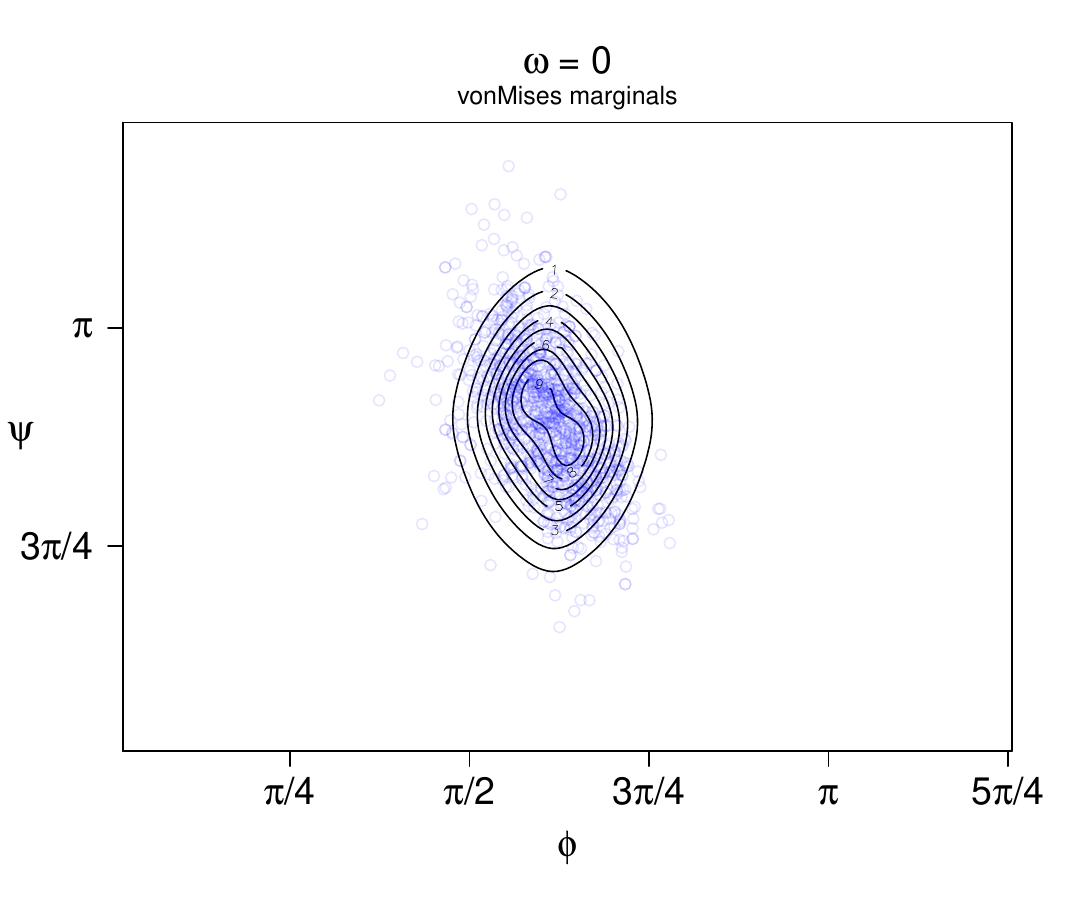} & 
        \includegraphics[width = 7.3cm, height = 7cm]{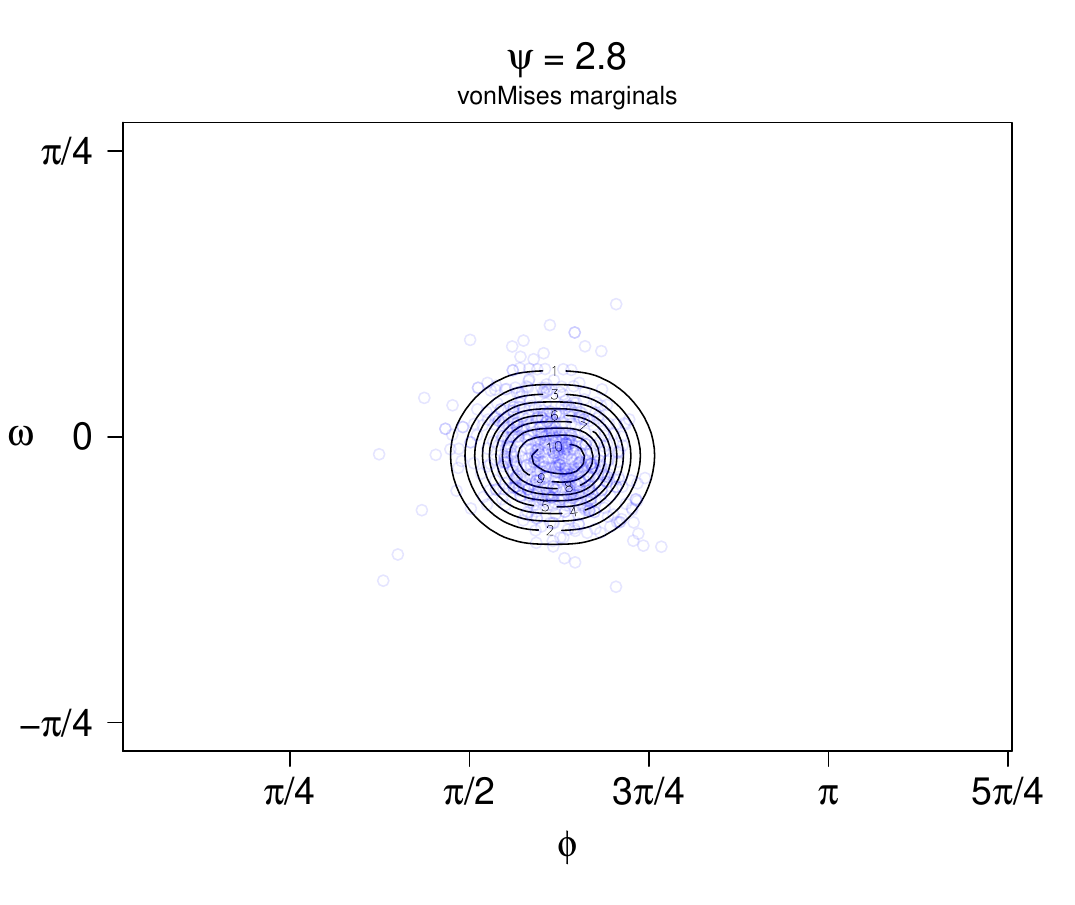} \\ 
        & \includegraphics[width = 7.3cm, height = 7cm]{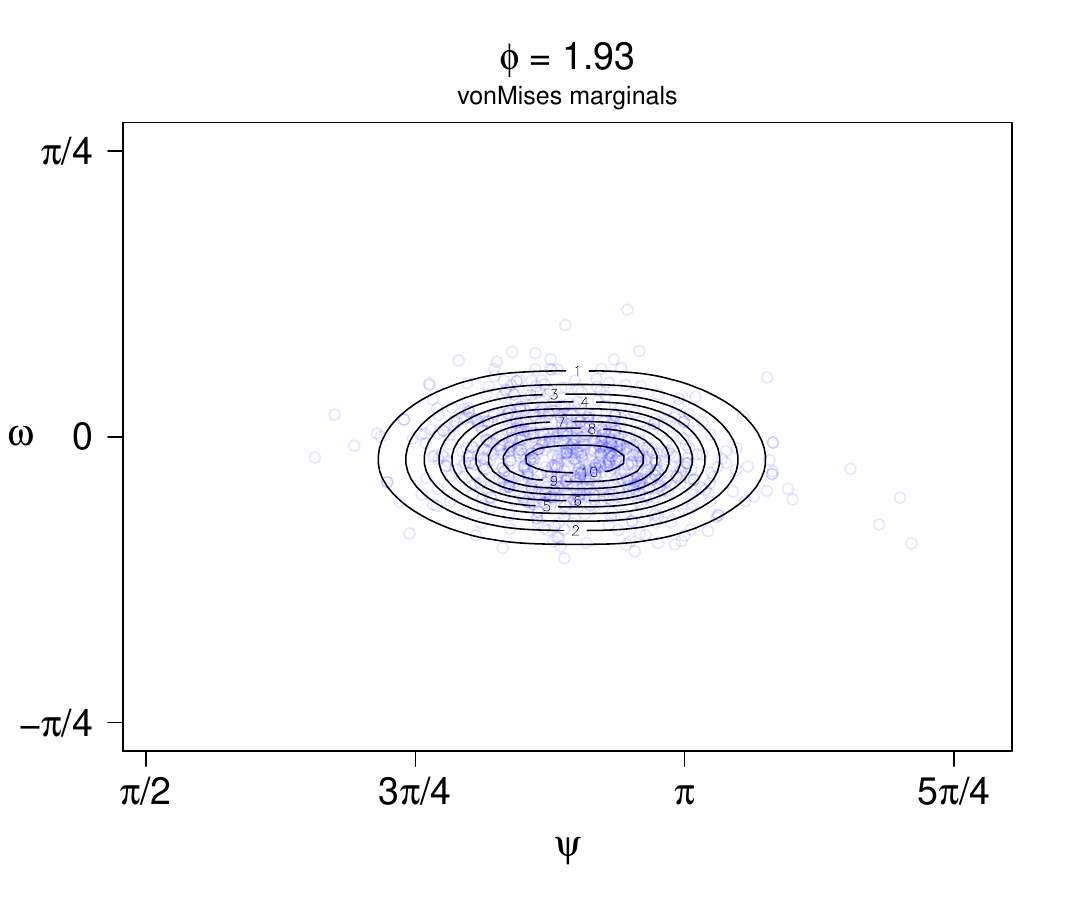}
\end{tabular}
\caption[]{Contour plots of TWCM
    with von Mises marginals, with parameter values estimated by maximum likelihood (IFM). The parameters of the marginals are $\hat{\mu}_1 = 1.93, \hat{\kappa}_1 = 27.6, \hat{\mu}_2 = 2.82, \hat{\kappa}_2 = 17.3, \hat{\mu}_3 = 6.23, \hat{\kappa}_3 = 84.4$, where $\hat{\mu}_i$ and $\hat{\kappa}_i$ denote the estimated value for $\mu$ and $\kappa$ of density \eqref{eq:dens_vonMises} corresponding to the marginal distribution of $\theta_i$ for $i\in\{1,2,3\}$, and the copula parameters are $\hat{\rho}_{12} = 9.18, \hat{\rho}_{13} = -1.17, \hat{\rho}_{23} = -0.09 $. \label{fig:protein_contour_plots} }
\end{figure}
}

\begin{table}
    \caption{\label{tab:competitors} Maximized log-likelihood, AIC, BIC, and the number of free parameters (denoted by $p$) \\  for the two models (TWCM and MNNTS)  for the protein dataset.}
    \centering
    \begin{tabular}{cccccc}
        \hline
        Model & Marginals & log-likelihood & AIC & BIC & $p$ \\
        \hline
         & uniform & -7920 & 15843 & 15855 & 2\\
          & wrapped Cauchy & 1131 & -2247 & -2202 & 8\\
         TWCM & cardioid & -3495 & 7005 & 7050 & 8\\
         & \textbf{von Mises} & \textbf{2046} & \textbf{-4076} & \textbf{-4031} & \textbf{8}\\
         & {Kato--Jones} & 1395 & -2762 & -2684 & 14\\
        \hline
        & ($M_1, M_2, M_3$) &&&& \\
        \hline
         & (0,0,0) & -11027 & 22055 & 22055 & 0\\
         & (1,1,1) & -6923 & 13874 & 13953 & 14\\
         & (2,2,2) & -4582 & 9268 & 9559 & 52\\
         & (3,3,3) & -2984 & 6220 & 6926 & 126\\
         MNNTS & (4,4,4) & -1811 & 4118 & 5507 & 248\\
         & (5,5,5) & -921 & 2702 & 5111 & 430\\
         & (6,6,6) & -231 & 1829 & 5660 & 684\\
         & (7,7,7) & 315 & 1414 & 7138 & 1022\\
         \hline
    \end{tabular}
\end{table}

We conclude this section by a comparison of our model with the MNNTS distribution~\eqref{MNNTS} of \cite{fernandez-duran_modeling_2014} {which we summarized in Section~\ref{sec:compet}}. Table~\ref{tab:competitors} presents the fit of various models.
The maximised log-likelihood, AIC and BIC are reported for each model, along with the number of free parameters, denoted by $p$.
The algorithms for fitting the MNNTS distribution are taken from the \textsc{R} package CircNNTSR \citep{CircNNTSR}.
Besides the reasons mentioned in Section~\ref{sec:compet} to compare our model to the MNNTS, it is also the only competitor for which we could find implemented and working code.
The number of free parameters for MNNTS($M_1, M_2, M_3$) is calculated as $2 \left(\prod_{i=1}^3 (M_i + 1) - 1 \right)$.
Various combinations of the values $M_1, M_2, M_3$ are shown in Table~\ref{tab:competitors}, with the number of free parameters increasing rapidly.
The MNNTS is not able to match the fit of our copula (not only in terms of AIC/BIC but even in terms of log-likelihood), even for a very large number of parameters.
The best model for all three measures of fit, the trivariate wrapped Cauchy copula with von Mises marginals, is shown in bold.

\subsection{Climate data}\label{sec:climpaper}
As discussed in the Introduction, the joint modeling of wave height and wave direction is a common approach in environmental studies aiming to identify sea regimes, with applications ranging from coastal erosion to offshore engineering and pollution tracking. Cylindrical distributions, such as the Abe–Ley distribution  \citep{abe_tractable_2017}, are typically employed for this purpose.
However, as previously emphasized, a more comprehensive analysis requires the inclusion of wind direction, which plays a critical role in shaping wave patterns, particularly in semi-enclosed basins. This is especially relevant in regions like the Adriatic Sea, where seasonal winds such as the Sirocco and Bora significantly influence wave dynamics \citep{lagona_hidden_2015}. 
In the following, we analyze a dataset that captures this triplet structure, that is, wave height (linear, $x$) and direction (circular, $\theta_1$), as well as wind direction (circular, $\theta_2$).
Such data thus is on a hyper-cylinder, which we can  analyze with our copula. 

For this purpose we use a time series of {1326 observations of} red half-hourly wave directions and heights as well as wind directions, recorded in the period 15/02/2010 – 16/03/2010 by the buoy of Ancona, located in the Adriatic Sea at about 30 km from the coast. {The data points are collected far spaced enough from each other to be considered independent.} 
We choose as our circular marginals the wrapped Cauchy distribution \eqref{eq:wc_density} and the Weibull distribution for the linear marginal, with density given by
\begin{equation*}
    f(x) = \frac{\nu}{\lambda} \left( \frac{x}{\lambda} \right)^{\nu-1} \exp^{- \left(x/\lambda \right)^\nu},
\end{equation*}
where $\nu$ is the shape parameter and $\lambda$ is the scale paramter.
Of course, many other combinations could be considered, but these work as we see. 
Our exploratory analysis shows that our data is multimodal and hence a mixture model is required, so we use densities of the form
\begin{equation} \label{eq:mixture_model}
    f(\theta_1, \theta_2, x) = \sum_{i=1}^K \pi_i f_i(\theta_1, \theta_2, x), \quad \sum_{i=1}^K \pi_i = 1,
\end{equation}
where $K$ is the number of components of the mixture model, $\pi_i$ is the weight of each class $i=1,\ldots, K$ and $f_i(\theta_1, \theta_2, x)$ is the density of component $i$ and is the copula as defined in \eqref{eq:extended_density}, with marginals as already explained.
In order to estimate the parameters, we use a variant of the Expectation-Maximization (EM) algorithm to find the values of the parameters for each component of the mixture model.
As with maximum likelihood estimation with non-uniform marginals, the maximisation is done as a first step for the parameters of the marginal distributions and then, using the obtained parameters, estimates of the parameters of the copula are obtained. The M-step of our algorithm thus adopts the IFM method described in Section~\ref{sec:adding marginals}, which is why we speak of a variant of the EM algorithm, which for the rest works exactly like an EM algorithm. The initial values for the parameters were randomly chosen.
This was repeated 10 times and the parameters that maximised the log-likelihood were chosen. To find the number of mixture components, we considered the values $K=2,3,4,5$ and used the BIC to determine the best-fitting model. We found that $K=4$ components fit best the data, {as the BIC values are 16045, 15850, {10750} and 15655, for $K=2,3,4,5$, respectively}. 

\begin{figure}[htbp]
    \centering
    \includegraphics[scale = 0.4]{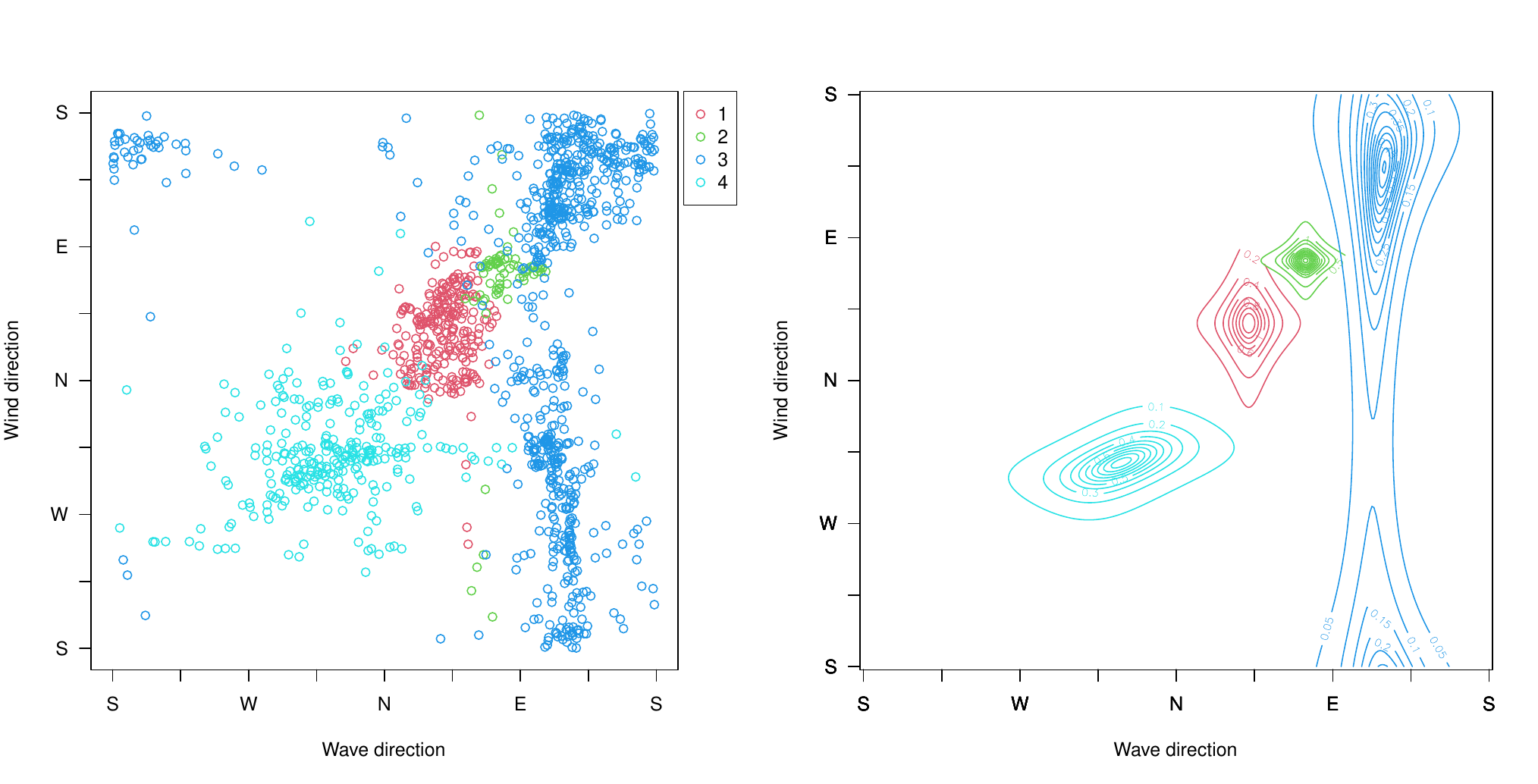}
    \includegraphics[scale = 0.4]{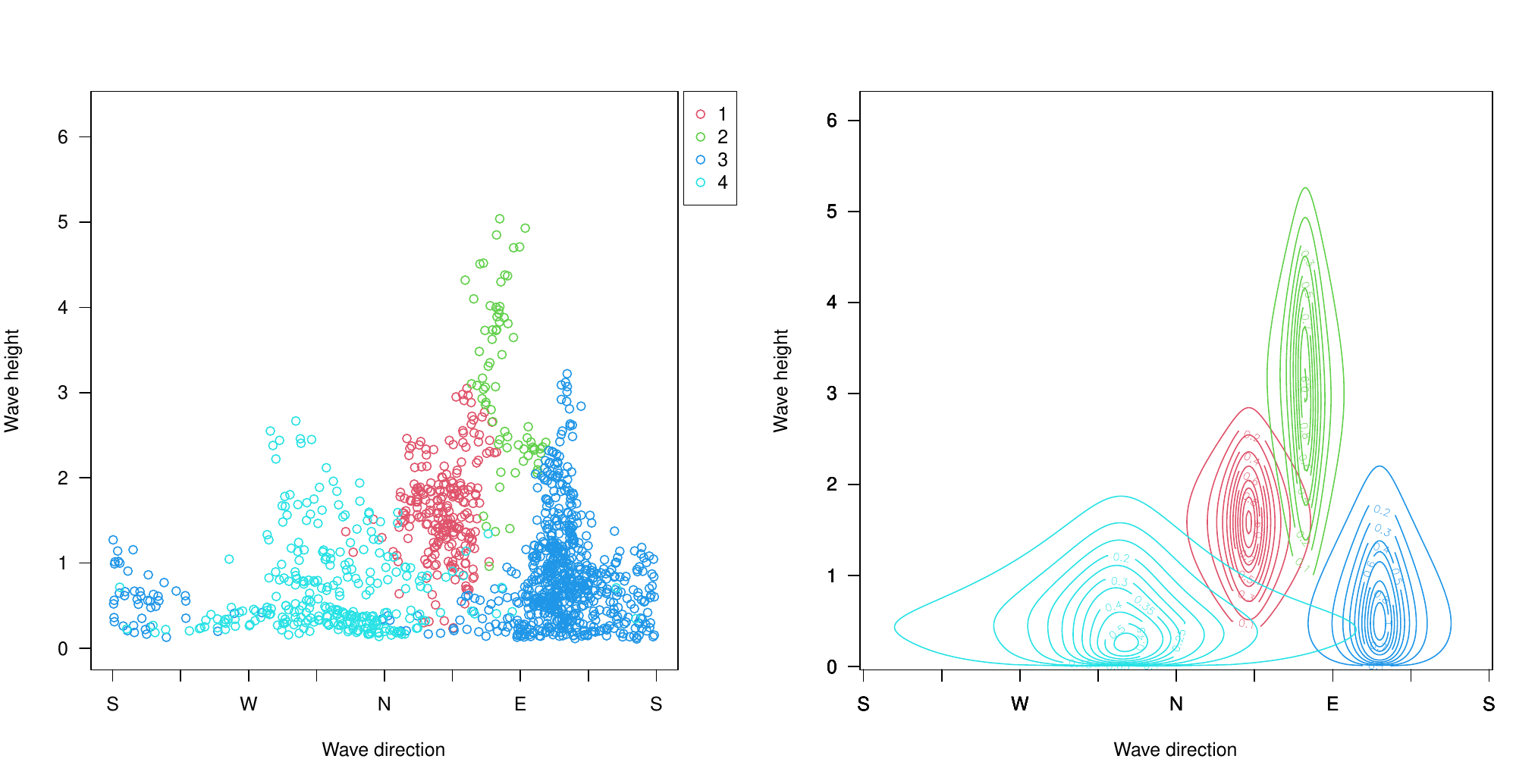}
    \includegraphics[scale = 0.4]{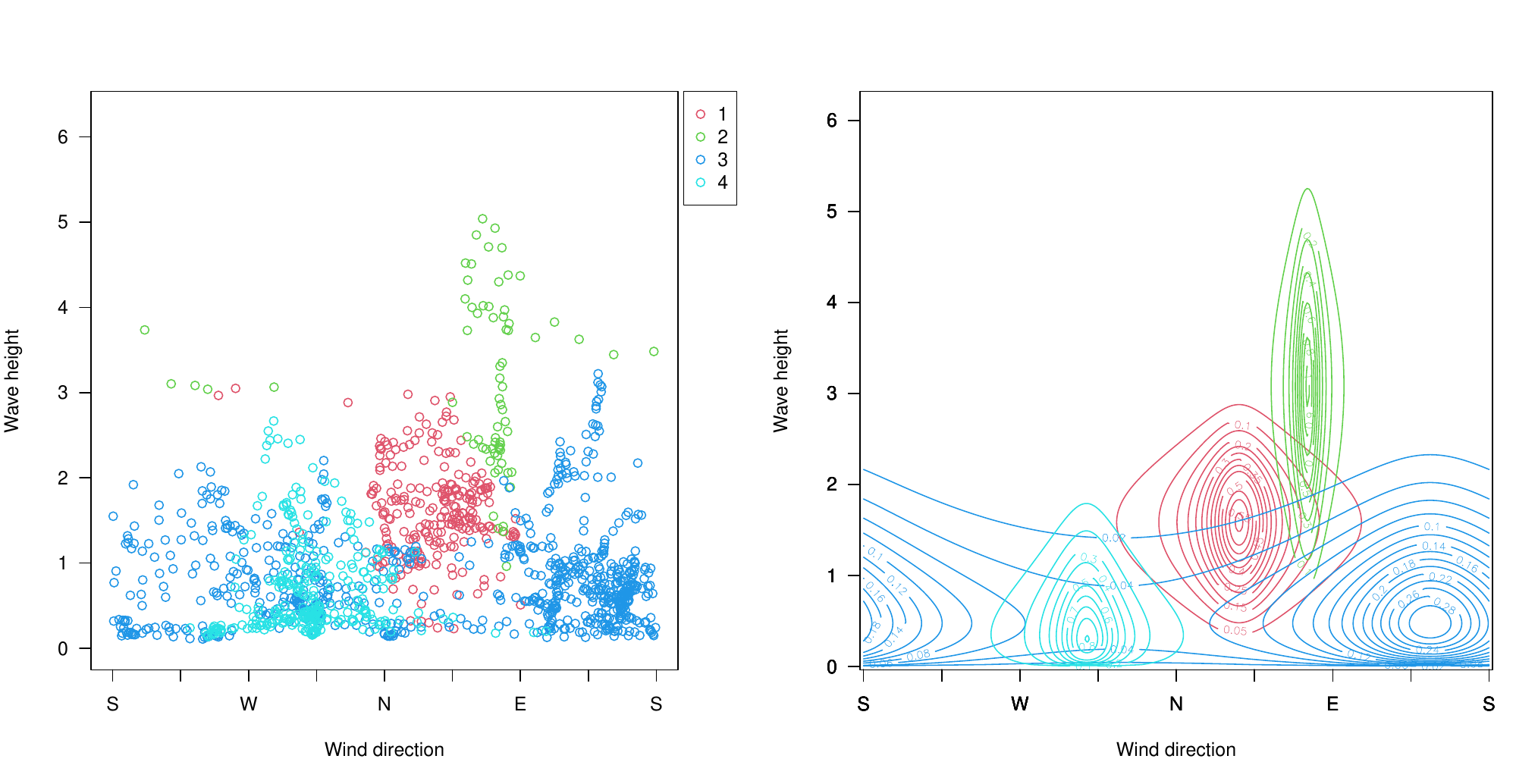}
    \caption{Contour plots for the climate data of the bivariate marginal distributions of our {extended TWCC} for the four different clusters as detected by our variant of the EM algorithm. Each cluster is presented in a different colour {and N, E, S, W represent north, east, south and west, respectively}.}
    \label{fig:wave_data}
\end{figure}

We now give the parameter estimates, plots of the data and clusters, and analyse the data based on our copula.
The parameter estimates for this mixture model are given in Table~\ref{tab:wave_data}. It should be noted that in each of the clusters, one of the $\rho_{ij}$ is large in absolute value, meaning that we are in the limit case discussed in \cite{paper1}. The values of the $\rho_{ij}$s are very different between each cluster, {with Cluster 3 having a very large negative value for $\hat{\rho}_{12}$, Cluster 1 having a smaller negative value for the same parameter and Cluster 2 having a large positive value for $\hat{\rho}_{13}$. The values of the parameters calculated for Cluster 4 are not as extreme.}
For visualization of the results, the bivariate marginal distribution is plotted in Figure~\ref{fig:wave_data}, as given in Theorem~\ref{theo_cop_mar}, with N, E, S, W representing north, east, south and west, respectively. The plots on the left hand side are scatter plots of the data, coloured according to their cluster as given by our variant of the EM algorithm with 4 components. The same colour is used for each cluster to plot the bivariate marginal distribution for each of the components.
{In the Adriatic sea, the predominant winds are bora, coming from north/north-east, and sirocco, from the south-east. The different winds are captured in different clusters of the data, with the red and green cluster being caused by the bora wind and dark blue by the sirocco wind.}

\begin{table}
    \caption{\label{tab:wave_data} Parameter estimates as obtained from our variant of the EM algorithm for 4 components. The marginal distributions for wind and wave directions are wrapped Cauchy, and the estimates of the parameters are denoted by $\hat{\xi}_{1}, \hat{\mu}_{1}$ and $\hat{\xi}_{2}, \hat{\mu}_{2}$, respectively, {with $\pi$ representing north.} For the wave height we use the Weibull distribution, where $\hat{\lambda}_{3}$ is the estimated scale parameter and $\hat{\nu}_{3}$  the estimated shape parameter. 
   Finally $\hat{\pi}_i$ denotes the estimate of the weight of  class $i$.}
    \centering
    \begin{tabular}{c|cccc}
        Parameters & Cluster 1 & Cluster 2 & Cluster 3 & Cluster 4 \\
        \hline
        $\hat{\rho}_{12}$ & -92.092 & -0.008 & -5956.914 & -5.168 \\ 
        $\hat{\rho}_{13}$ & 43.331 & 584.384 & -0.032 & 0.864 \\ 
        $\hat{\rho}_{23}$ & -0.0003 & -0.218 & 0.005 & -0.224 \\ 
        $\hat{\mu}_{1}$ & 3.869 & 4.440 & 5.184 & 2.588 \\ 
        $\hat{\xi}_{1}$ & 0.834 & 0.882 & 0.809 & 0.578 \\ 
        $\hat{\mu}_{2}$ & 3.772 & 4.460 & 5.692 & 2.239 \\ 
        $\hat{\xi}_{2}$ & 0.729 & 0.887 & 0.432 & 0.736 \\ 
        $\hat{\lambda}_{3}$ & 1.781 & 3.414 & 0.948 & 0.766 \\ 
        $\hat{\nu}_{3}$ & 3.207 & 3.390 & 1.531 & 1.424 \\ 
        $\hat{\pi}_i$ & 0.184 & 0.054 & 0.546 & 0.216 \\ 
    \end{tabular}
\end{table}

\section{Future research} \label{sec:conclusion}

{The flexibility of the TWCM allows for several applications of the distribution, as well as further extensions.
Regarding protein structure prediction, the methodology of the seminal paper \cite{boomsma2008generative} uses a trivariate distribution with $\omega$ as discrete variable in their software TorusDBN and in the future it would be worth investigating if using the TWCM instead of their trivariate distribution leads to any improvement in prediction.}

We now discuss a few extensions.
First, we might wish to introduce asymmetry into our copula {TWCC($\boldsymbol{\rho}$)}. Skewness can be handled at the level of the marginals via suitably choosing skew marginal distributions, but this does not allow altering the symmetry of the dependence structure, hence the copula itself. For example, this  can be done by adopting the approach of \cite{AL22} which consists in multiplying \eqref{eq:tri_density} with the skewing function $(1+\lambda_1\sin(u_1)+\lambda_2\sin(u_2)+\lambda_3\sin(u_3))$ for $\lambda_1,\lambda_2,\lambda_3\in(-1,1)$ satisfying $|\lambda_1|+|\lambda_2|+|\lambda_3|<1$. When all three skewness parameters are 0, we retrieve the original copula, and as soon as one parameter deviates from zero, we obtain a skew version of  our {TWCC($\boldsymbol{\rho}$)} given by \eqref{eq:tri_density}. As future work, it will be interesting to see in how far asymmetry in the copula can add flexibility on top of asymmetry in the marginals, and how both can be ideally combined.

Second, in a similar manner as in \cite{WJ80}, the TWCM can be used to construct an AR(2) process on the circle.
Let $\Theta_0, \Theta_1, \ldots,$ be $[0,2 \pi)$-valued random variables on the circle such that
\begin{align}
p(\theta_0,\theta_1) & = f(\theta_0,\theta_1), \nonumber \\
p(\theta_t | \theta_{t-1},\ldots,\theta_0) & = p (\theta_t | \theta_{t-1},\theta_{t-2}) =  \frac{ | 1-\delta_t^2 | }{1+\delta_t^2 - 2 \delta_t \cos ( 2 \pi F(\theta_t)- \eta_t)} f(\theta_t), \quad t=2,3,\ldots, \label{eq:transition_density}
\end{align}
{where $p(\theta_0,\theta_1)$ is the initial distribution and $p(\theta_t|\theta_{t-1},\theta_{t-2})$ is the transition density.
Here $f(\theta_t)$ is a density on the circle $[0,2\pi)$, $F(\theta_t) = \int_{c}^{\theta_t} f(x) dx $ for some arbitrary origin $c$ on the circle, and $f(\theta_0,\theta_1)$ is a density on the torus $[0,2\pi)^2$ with a common univariate marginal density $f$.
Also, the parameters are assumed to satisfy $\eta_t =  \arg (\phi_t),$ $ \delta_t =|\phi_t|,$ $\phi_t = - \rho_{t-1,t-2} \{ \rho_{t,t-1}^{-1} e^{2 \pi {\rm i} F(\theta_{t-1})} + \rho_{t,t-2 }^{-1} e^{2 \pi {\rm i}  F(\theta_{t-2})}\}$, $\rho_{t,t-1}, \rho_{t,t-2}, \rho_{t-1,t-2} \in \mathbb{R}$, $\rho_{t,t-1} \cdot \rho_{t,t-2} \cdot \rho_{t-1,t-2} >0$, and $|\rho_{j k}| < |\rho_{ij} \rho_{i k}| / ( |\rho_{ij}| + |\rho_{i k}|)$ for $(i,j,k)$ a permutation of $(t,t-1,t-2)$.
{The transition density (\ref{eq:transition_density}) is derived by substituting $(\theta_t,\theta_{t-1},\theta_{t-2})$ into $(\theta_1,\theta_2,\theta_3)$ in (\ref{eq:extended_density}) with the common marginal density $f$ and calculating its univariate conditional density.
Hence, it is clear that this circular AR(2) process $\{\Theta_i\}$, under the conditions above, is strictly stationary in the sense that $(\Theta_{t_1+\tau}, \ldots, \Theta_{t_n+\tau}) \overset{\mathrm{d}}{=} (\Theta_{t_1}, \ldots, \Theta_{t_n})$ for any nonnegative integers $t_1, \ldots, t_n, \tau$ and any positive integer $n$.}}
If $f$ is the wrapped Cauchy density (\ref{eq:wc_density}), then the transition density (\ref{eq:transition_density}) can also be expressed in closed form without integrals.  The tractability  of our model would  make {the circular AR(2) process} very appealing and important for time-dependent directional data.

We also  mention that a general $d$-dimensional extension of the TWCM follows {in the same way as} the $d$-dimensional generalization of the TWCC  in \cite{paper1} by again extending the uniform marginals to any marginals. Moreover, as alternative to the TWCM one could estimate the marginals  in a non-parametric way and then combine the estimated marginals with our TWCC  (see {\cite{genest_semiparametric_1995}} for such a procedure in $\R^d$).

\textbf{Code availability:} The relevant code of the paper is available in \url{https://doi.org/10.5281/zenodo.15675162} as well as in \url{https://github.com/Sophia-Loizidou/Trivariate-wrapped-Cauchy-copula}.

\textbf{Acknowledgements:} The authors would like to thank Thomas Hamelryck and Ola R\o nning for the protein data. Kanti Mardia would like to thank the Leverhulme Trust for the Emeritus Fellowship.

\textbf{Funding:} Shogo Kato was supported by JSPS KAKENHI Grant Number 20K03759 and 25K15037.
Sophia Loizidou was supported by the grant PRIDE/21/16747448/MATHCODA from the Luxembourg National Research Fund.


\bibliographystyle{abbrv}
\bibliography{references_multcop}  

@preamble{" \newcommand{\noop}[1]{} "}

@Book{Joe15,
  Title                    = {Dependence Modeling with Copulas},
  Author                   = {Joe, Harry},
  Publisher                = {CRC Press},
  Year                     = {2015},
  Address                  = {Boca Raton, FL, USA}
}

@article{mardia2025fisher,
  title={Fisher’s legacy of directional statistics, and beyond to statistics on manifolds},
  author={Mardia, Kanti V},
  journal={Journal of Multivariate Analysis},
  volume={207},
  pages={105404},
  year={2025},
  publisher={Elsevier}
}

@article{lane2023protein,
  title={Protein structure prediction has reached the single-structure frontier},
  author={Lane, Thomas J},
  journal={Nature Methods},
  volume={20},
  number={2},
  pages={170--173},
  year={2023},
  publisher={Nature Publishing Group US New York}
}

@article{NFT2017,
	author = {Navarro, A.K.W. and Frellsen, J. and Turner, R.E.},
	journal = {Proceedings of the Thirty-First AAAI Conference on Artificial Intelligence (AAAI-17)},
	pages = {2394--2400},
	title = {{The multivariate generalised von Mises distribution: inference and applications}},
	year = {2017}}

@article{Selvitella2019,
author = {Alessandro Selvitella},
title = {{On geometric probability distributions on the torus with applications to molecular biology}},
volume = {13},
journal = {Electronic Journal of Statistics},
number = {2},
publisher = {Institute of Mathematical Statistics and Bernoulli Society},
pages = {2717 -- 2763},
year = {2019},
doi = {10.1214/19-EJS1579},
URL = {https://doi.org/10.1214/19-EJS1579}
}

@article{AL22,
	author = {Ameijeiras-Alonso, Jose and Ley, Christophe},
	journal = {Biostatistics},
	pages = {685--704},
	title = {Sine-skewed toroidal distributions and their application in protein bioinformatics},
	volume = {23},
	year = {2022}}

@book{faltinsen1993sea,
  title={Sea loads on ships and offshore structures},
  author={Faltinsen, Odd},
  volume={1},
  year={1993},
  publisher={Cambridge University Press}
}

@article{pleskachevsky2009interaction,
  title={Interaction of waves, currents and tides, and wave-energy impact on the beach area of {Sylt Island}},
  author={Pleskachevsky, Andrey and Eppel, Dieter P and Kapitza, Hartmut},
  journal={Ocean Dynamics},
  volume={59},
  number={3},
  pages={451--461},
  year={2009},
  publisher={Springer}
}

@article{huang2011wave,
  title={Wave-induced drift of small floating objects in regular waves},
  author={Huang, Guoxing and Law, Adrian Wing-Keung and Huang, Zhenhua},
  journal={Ocean Engineering},
  volume={38},
  number={4},
  pages={712--718},
  year={2011},
  publisher={Elsevier}
}

@article{WJ80,
	author = {Wehrly, T. E. and Johnson, R. A.},
	journal = {Biometrika},
	pages = {255-256},
	title = {Bivariate models for dependence of angular observations and a related {M}arkov process},
	volume = {66},
	year = {1980}}

@article{Mardia13,
	author = {Mardia, K. V.},
	journal = {Journal of the Royal Statistical Society: Series C (Applied Statistics)},
	pages = {487--514},
	title = {Statistical approaches to three key challenges in protein structural bioinformatics},
	volume = {62},
	year = {2013}}

@book{LV19,
	address = {Boca Rat{\'o}n, Florida},
	editor = {Ley, C. and Verdebout, T.},
	publisher = {Chapman and Hall/CRC Press},
	title = {Applied Directional Statistics: Modern Methods and Case Studies},
	year = {2018}}

@article{HBPFJH10,
	author = {Harder, T. and Boomsma, W. and Paluszewski, M. and Frellsen, Jes and Johansson, K. E. and Hamelryck, T.},
	journal = {BMC Bioinformatics},
	pages = {306},
	title = {Beyond rotamers: a generative, probabilistic model of side chains in proteins},
	volume = {11},
	year = {2010}}

@article{GSMH19,
	author = {Garc\'ia-Portugu\'es, E. and S\"orensen, M. and Mardia, Kanti V and Hamelryck, T.},
	journal = {Statistics and Computing},
	pages = {1--22},
	title = {Langevin diffusions on the torus: estimation and applications},
	volume = {29},
	year = {2019}}

@article{BMTFKH08,
	author = {Boomsma, W. and Mardia, Kanti V and Taylor, Charles C and Ferkinghoff-Borg, J. and Krogh, A. and Hamelryck, T.},
	journal = {Proceedings of the National Academy of Sciences},
	pages = {8932--8937},
	title = {A generative, probabilistic model of local protein structure},
	volume = {105},
	year = {2008}}

@book{HMF12,
	address = {Heidelberg, Germany},
	editor = {Hamelryck, T. and Mardia, Kanti V and Ferkinghoff-Borg, J.},
	publisher = {Springer},
	title = {Bayesian {M}ethods in {S}tructural {B}ioinformatics},
	year = {2012}}

@article{circulas,
	author = {Jones, M. C. and Pewsey, Arthur and Kato, Shogo},
	journal = {Annals of the Institute of Statistical Mathematics},
	pages = {843--862},
	title = {On a class of circulas: copulas for circular distributions},
	volume = {67},
	year = {2015}}

@article{Baba1981,
	author = {Baba, Y},
	journal = {Proceedings of the Institute of Statistical Mathematics},
	pages = {41--54},
	title = {Statistics of angular data: wrapped normal distribution model (in Japanese)},
	volume = {28},
	year = {1981}}

@article{mardia2008,
	author = {Mardia, Kanti V and Hughes, Gareth and Taylor, Charles C and Singh, Harshinder},
	journal = {Canadian Journal of Statistics},
	pages = {99--109},
	publisher = {Wiley Online Library},
	title = {A multivariate von {M}ises distribution with applications to bioinformatics},
	volume = {36},
	year = {2008}}

@article{mardia2007,
	author = {Mardia, Kanti V and Taylor, Charles C and Subramaniam, Ganesh K},
	journal = {Biometrics},
	pages = {505--512},
	publisher = {Wiley Online Library},
	title = {Protein bioinformatics and mixtures of bivariate von {M}ises distributions for angular data},
	volume = {63},
	year = {2007}}

@article{kP15,
    author = {Kato, Shogo and Pewsey, Arthur},
    title = "{A Möbius transformation-induced distribution on the torus}",
    journal = {Biometrika},
    volume = {102},
    number = {2},
    pages = {359-370},
    year = {2015},
    month = {03},
}

@book{mardia2000,
	address = {Chichester, United Kingdom},
	author = {Mardia, Kanti V and Jupp, Peter E},
	publisher = {John Wiley \& Sons},
	title = {Directional Statistics},
	year = {2000}}

@article{paper1,
  title={The trivariate wrapped Cauchy copula},
  author={Kato, Shogo and Ley, Christophe and Loizidou, Sophia and Mardia, Kanti V.},
  journal={arXiv},
  volume = {2401.10824},
  year={2025},
  url={https://arxiv.org/abs/2401.10824},
  doi={10.48550/arXiv.2401.10824}
}

@book{joe_multivariate_1997,
	location = {New York},
	title = {Multivariate Models and Multivariate Dependence Concepts},
	isbn = {978-0-367-80389-6},
	pagetotal = {424},
	publisher = {Chapman and Hall/{CRC}},
	author = {Joe, Harry},
	date = {1997-05-01},
    year = {1997},
	doi = {10.1201/9780367803896},
}

@article{cock_biopython_2009,
	title = {Biopython: freely available {Python} tools for computational molecular biology and bioinformatics},
	volume = {25},
	issn = {1367-4811},
	shorttitle = {Biopython},
	language = {eng},
	number = {11},
	journal = {Bioinformatics (Oxford, England)},
	author = {Cock, Peter J. A. and Antao, Tiago and Chang, Jeffrey T. and Chapman, Brad A. and Cox, Cymon J. and Dalke, Andrew and others},
	year = {2009},
	pmid = {19304878},
	pmcid = {PMC2682512},
	pages = {1422--1423},
}

@article{kabsch_dictionary_1983,
	title = {Dictionary of protein secondary structure: {Pattern} recognition of hydrogen-bonded and geometrical features},
	volume = {22},
	copyright = {Copyright © 1983 John Wiley \& Sons, Inc.},
	issn = {1097-0282},
	shorttitle = {Dictionary of protein secondary structure},
	language = {en},
	number = {12},
	urldate = {2023-11-07},
	journal = {Biopolymers},
	author = {Kabsch, Wolfgang and Sander, Christian},
	year = {1983},
	pages = {2577--2637},
}

@article{jacobsen_introduction_2023,
	title = {Introduction to {Protein} {Structure}},
	url = {http://arxiv.org/abs/2307.02169},
	doi = {10.48550/arXiv.2307.02169},
	urldate = {2023-10-12},
	publisher = {arXiv},
	author = {Jacobsen, Annika and van Dijk, Erik and Mouhib, Halima and Stringer, Bas and Ivanova, Olga and Gavaldá-Garciá, Jose and others},
	year = {2023},
    journal = {arXiv},
    volume = {2307.02169},
}

@book{jammalamadaka_topics_2001,
	title = {Topics in Circular Statistics},
	isbn = {978-981-277-926-7},
	pagetotal = {348},
	publisher = {World Scientific},
	author = {Jammalamadaka, S. Rao and Sengupta, Ashis},
	year = {2001},
	langid = {english},
}

@incollection{mardia_directional_2018,
	title = {Directional Distributions},
	rights = {Copyright © 2018 John Wiley \& Sons, Ltd. All rights reserved.},
	isbn = {978-1-118-44511-2},
	pages = {1--13},
	booktitle = {Wiley {StatsRef}: Statistics Reference Online},
	publisher = {John Wiley \& Sons, Ltd},
	author = {Mardia, Kanti V. and Ley, Christophe},
	urldate = {2023-11-13},
	year = {2018},
	langid = {english},
}

@article{pewsey_recent_2021,
	title = {Recent advances in directional statistics (with discussion)},
	volume = {30},
	issn = {1863-8260},
	pages = {1--58},
	number = {1},
	journal = {{TEST}},
	shortjournal = {{TEST}},
	author = {Pewsey, Arthur and García-Portugués, Eduardo},
	urldate = {2023-11-13},
	date = {2021-03-01},
    year = {2021},
	langid = {english},
}

@article{johnson_angular-linear_1978,
	title = {Some Angular-Linear Distributions and Related Regression Models},
	volume = {73},
	issn = {0162-1459},
	pages = {602--606},
	number = {363},
	journal = {Journal of the American Statistical Association},
	author = {Johnson, Richard A. and Wehrly, Thomas E.},
	urldate = {2023-11-13},
	date = {1978-09-01},
    year = {1978},
}

@article{abe_tractable_2017,
	title = {A tractable, parsimonious and flexible model for cylindrical data, with applications},
	volume = {4},
	issn = {2452-3062},
	pages = {91--104},
    journal = {Econometrics and Statistics},
	journaltitle = {Econometrics and Statistics},
	shortjournal = {Econometrics and Statistics},
	author = {Abe, Toshihiro and Ley, Christophe},
	urldate = {2023-11-13},
	date = {2017-10-01},
    year = {2017},
}

@article{lagona_hidden_2015,
	title = {A hidden {M}arkov model for the analysis of cylindrical time series},
	volume = {26},
	rights = {Copyright © 2015 John Wiley \& Sons, Ltd.},
	issn = {1099-095X},
	url = {https://onlinelibrary.wiley.com/doi/abs/10.1002/env.2355},
	doi = {10.1002/env.2355},
	pages = {534--544},
	number = {8},
	journal = {Environmetrics},
	author = {Lagona, Francesco and Picone, Marco and Maruotti, Antonello},
	urldate = {2023-11-13},
	year = {2015},
	langid = {english},
}

@article{mastrantonio_joint_2018,
	title = {The joint projected normal and skew-normal: A distribution for poly-cylindrical data},
	volume = {165},
	issn = {0047-259X},
	url = {https://www.sciencedirect.com/science/article/pii/S0047259X17301069},
	doi = {10.1016/j.jmva.2017.11.006},
	shorttitle = {The joint projected normal and skew-normal},
	pages = {14--26},
	journal = {Journal of Multivariate Analysis},
	shortjournal = {Journal of Multivariate Analysis},
	author = {Mastrantonio, Gianluca},
	urldate = {2023-11-08},
	date = {2018-05-01},
    year = {2018},
}

@article{senior2019protein,
  title={Protein structure prediction using multiple deep neural networks in the 13th {Critical Assessment of Protein Structure Prediction (CASP13)}},
  author={Senior, Andrew W and Evans, Richard and Jumper, John and Kirkpatrick, James and Sifre, Laurent and Green, Tim and others},
  journal={Proteins: Structure, Function, and Bioinformatics},
  volume={87},
  number={12},
  pages={1141--1148},
  year={2019},
  publisher={Wiley Online Library}
}

@article{fernandez-duran_modeling_2014,
	title = {Modeling angles in proteins and circular genomes using multivariate angular distributions based on multiple nonnegative trigonometric sums},
	volume = {13},
	rights = {De Gruyter expressly reserves the right to use all content for commercial text and data mining within the meaning of Section 44b of the German Copyright Act.},
	issn = {1544-6115},
	url = {https://www.degruyter.com/document/doi/10.1515/sagmb-2012-0012/html},
	doi = {10.1515/sagmb-2012-0012},
	pages = {1--18},
	number = {1},
	journal = {Statistical Applications in Genetics and Molecular Biology},
	author = {Fern\'andez-Dur\'an, Juan Jos\'e and Gregorio-Dom\'inguez, {MarÍa} Mercedes},
	urldate = {2023-11-08},
	date = {2014-02-01},
    year = {2014},
	langid = {english},
}

@Article{CircNNTSR,
    title = {{CircNNTSR}: An {R} Package for the Statistical Analysis of Circular, Multivariate Circular, and Spherical Data Using Nonnegative Trigonometric Sums},
    author = {Juan Jose Fern\'andez-Dur\'an and Maria Mercedes Gregorio-Dom\'inguez},
    journal = {Journal of Statistical Software},
    year = {2016},
    volume = {70},
    number = {6},
    pages = {1--19},
    doi = {10.18637/jss.v070.i06},
  }

@article{kim_multivariate_2016,
	title = {A multivariate circular distribution with applications to the protein structure prediction problem},
	volume = {143},
	issn = {0047-259X},
	url = {https://www.sciencedirect.com/science/article/pii/S0047259X1500250X},
	doi = {10.1016/j.jmva.2015.09.024},
	pages = {374--382},
	journaltitle = {Journal of Multivariate Analysis},
    journal = {Journal of Multivariate Analysis},
	shortjournal = {Journal of Multivariate Analysis},
	author = {Kim, Sungsu and {SenGupta}, Ashis and Arnold, Barry C.},
	urldate = {2023-12-01},
	date = {2016-01-01},
    year = {2016},
}

@misc{genna_sars-cov-2_2020,
	title = {{SARS}-{CoV}-2 {Inhibition}, {Host} {Selection} and {Next}-{Move} {Prediction} {Through} {High}-{Performance} {Computing}},
	urldate = {2023-10-17},
	author = {Genna, Vito and Hospital, Adam and Orozco, Modesto},
	year = {2020},
}

@article{genest_semiparametric_1995,
	title = {A semiparametric estimation procedure of dependence parameters in multivariate families of distributions},
	volume = {82},
	issn = {0006-3444},
	url = {https://doi.org/10.1093/biomet/82.3.543},
	doi = {10.1093/biomet/82.3.543},
	number = {3},
	journal = {Biometrika},
	author = {Genest, C. and Ghoudi, K. and Rivest, L.-P.},
	year = {1995},
	pages = {543--552},
}

@inproceedings{thygesen_efficient_2021,
	title = {Efficient {Generative} {Modelling} of {Protein} {Structure} {Fragments} using a {Deep} {Markov} {Model}},
	url = {https://proceedings.mlr.press/v139/thygesen21a.html},
	language = {en},
	urldate = {2024-01-13},
	booktitle = {Proceedings of the 38th {International} {Conference} on {Machine} {Learning}},
	publisher = {PMLR},
	author = {Thygesen, Christian B. and Steenmans, Christian Skjødt and Al-Sibahi, Ahmad Salim and Moreta, Lys Sanz and Sørensen, Anders Bundgård and Hamelryck, Thomas},
	year = {2021},
	pages = {10258--10267},
}

@article{berkholz2012nonplanar,
  title={Nonplanar peptide bonds in proteins are common and conserved but not biased toward active sites},
  author={Berkholz, Donald S and Driggers, Camden M and Shapovalov, Maxim V and Dunbrack Jr, Roland L and Karplus, P Andrew},
  journal={Proceedings of the National Academy of Sciences},
  volume={109},
  number={2},
  pages={449--453},
  year={2012},
  publisher={National Acad Sciences},
    doi = {10.1073/pnas.1107115108},
}

@incollection{Mardiachapter,
	address = {Singapore},
	title = {Mixture {Models} for {Spherical} {Data} with {Applications} to {Protein} {Bioinformatics}},
	isbn = {978-981-19104-4-9},
	url = {https://doi.org/10.1007/978-981-19-1044-9\_2},
	language = {en},
	urldate = {2024-06-06},
	booktitle = {Directional {Statistics} for {Innovative} {Applications}: {A} {Bicentennial} {Tribute} to {Florence} {Nightingale}},
	publisher = {Springer Nature},
	author = {Mardia, Kanti V. and Barber, Stuart and Burdett, Philippa M. and Kent, John T. and Hamelryck, Thomas},
	editor = {SenGupta, Ashis and Arnold, Barry C.},
	year = {2022},
	doi = {10.1007/978-981-19-1044-9\_2},
	pages = {15--32},
}

@article{boomsma2008generative,
  title={A generative, probabilistic model of local protein structure},
  author={Boomsma, Wouter and Mardia, Kanti V and Taylor, Charles C and Ferkinghoff-Borg, Jesper and Krogh, Anders and Hamelryck, Thomas},
  journal={Proceedings of the National Academy of Sciences},
  volume={105},
  number={26},
  pages={8932--8937},
  year={2008},
  publisher={National Academy of Sciences}
}

\end{document}